\documentclass[
aps,
pra, 
twocolumn,
amsmath,amssymb,
longbibliography,
superscriptaddress]{revtex4-1}

\usepackage{graphicx,siunitx}
\usepackage[utf8]{inputenc} 
\usepackage[english]{babel}
\usepackage{hyperref}
\hypersetup{
colorlinks=true,
linkcolor=blue,
citecolor=blue,
urlcolor=blue}

\newcommand{\singlequote}[1]{`#1'}

\newcommand{\nii}[1]{_{\mathrm{#1}}} 
\newcommand{\rmd}{\mathrm{d}}

\newcommand{\SInsp}[2]{\SI[number-unit-product={}]{#1}{#2}} 
\newcommand{\ppmw}{\percent/\milli\watt} 

\newcommand{\todo}[1]{}

\begin{document}


\title{More efficient second harmonic generation of whispering gallery modes \\ by selective out-coupling}


\author{Luke S. Trainor}
\affiliation{The Dodd-Walls Centre for Photonic and Quantum Technologies, Department of Physics, University of Otago, 730 Cumberland Street, Dunedin 9016, New Zealand}
\author{Florian Sedlmeir}
\affiliation{The Dodd-Walls Centre for Photonic and Quantum Technologies, Department of Physics, University of Otago, 730 Cumberland Street, Dunedin 9016, New Zealand}
\affiliation{Max Planck Institute for the Science of Light, Staudtstra{\ss}e 2, 90158 Erlangen, Germany}
\author{Christian Peuntinger}
\author{Harald G. L. Schwefel}
\email[]{harald.schwefel@otago.ac.nz}
\affiliation{The Dodd-Walls Centre for Photonic and Quantum Technologies, Department of Physics, University of Otago, 730 Cumberland Street, Dunedin 9016, New Zealand}


\date{\today}

\begin{abstract}
We demonstrate second harmonic generation (SHG) in an $x$-cut congruent lithium niobate (LN) whispering gallery mode resonator.
We first show theoretically that independent control of the coupling of the pump and signal modes is optimal for high conversion rates.
A scheme based on our earlier work in Ref.\,\cite{sedlmeir_polarization-selective_2017} is then implemented experimentally to verify this.
Thereby we are able to improve on the efficiency of SHG by more than an order of magnitude by selectively out-coupling using a LN prism, utilizing the birefringence of it and the resonator in kind.
We report \SInsp{5.28}{\ppmw} efficiency for SHG from \SI{1555.4}{\nano\meter} to \SI{777.7}{\nano\meter}.
\end{abstract}


\maketitle


\section{Introduction \label{Introduction}}

Second harmonic generation (SHG) was one of the first nonlinear processes to be observed using the invention of the laser \cite{franken_generation_1961}.
It has found wide use in the frequency doubling of lasers and makes hard to reach frequency domains more accessible.
However, SHG is dictated by the weak second-order nonlinear susceptibility, and hence usually requires high-power pump lasers.
In order to increase the process's efficiency a cavity can be used, e.g. by placing the nonlinear crystal inside a bow-tie resonator.
Such a method can be very efficient, but requires specially coated mirrors which limit the bandwidth of the operation and are bulky and expensive to set up.
Recent monolithic implementations \cite{furst_naturally_2010,furst_second-harmonic_2015} achieved high conversion efficiencies by resonantly confining the interacting light fields within a whispering gallery mode (WGM) resonator \cite{sturman_generic_2011}.

WGM resonators are natural vessels to study nonlinear effects due to an unparalleled combination of favorable characteristics \cite{strekalov_nonlinear_2016,breunig_three-wave_2016}. 
The method of entrapment is total internal reflection, which allows for very good spatial confinement along the rim of the resonator.
Furthermore, total internal reflection is only very weakly dependent on the wavelength of light, which allows a large range of wavelengths to be simultaneously resonant in the same cavity; the range depends solely on the absorbency of the bulk dielectric.
This allows for resonances with very high $Q$ factors and finesse, greatly enhancing the contained fields, and hence the nonlinear effects of the bulk medium.
The efficiency is high as both the pump and the harmonic light can be resonantly enhanced and propagate with the same phase velocity within the WGM resonator.
This phase matching can be achieved in some birefringent nonlinear crystals and is highly dependent on the exact dispersion of the crystal.
By chance of nature some crystals allow natural phase matching for SHG, albeit in very limited frequency domains \cite{furst_naturally_2010,furst_second-harmonic_2015}.
Slightly broader operation is afforded through periodic poling of the crystal \cite{beckmann_highly_2011}.
An elegant method for broadband operation is to use the WGM in a distinct crystal orientation, namely the $x$ cut (or $y$ cut), where the optic axis of a uniaxial birefingent crystal lies within the plane of the equator \cite{furst_whispering_2016}.
In such a system the refractive index of the transverse magnetic (TM) mode---in addition to the effective nonlinear susceptibility---oscillates as the light propagates around the resonator's rim.
This allows broadband phase matching at slightly reduced efficiency \cite{lin_wide-range_2013,lin_continuous_2014}.
Although such WGM implementations still can be highly efficient, they are intrinsically limited by the external coupling rate of the harmonic light. 
Coupling to WGMs is usually achieved by overlapping the evanescent field of the resonator with the evanescent field of, e.g. a prism.
Therefore, typically either the pump is over-coupled or the signal is under-coupled, as the evanescent field decay length scales with wavelength.
The maximum of prism out-coupled harmonic power does not coincide with the maximum intrinsic harmonic light.
By exploiting the birefringence of the WGM and the coupling prism it is however possible to independently control the coupling rates for pump and second harmonic modes, and thereby selectively out-couple the light.

In a previous work, we developed the theoretical model for selective coupling and demonstrated it in the linear regime \cite{sedlmeir_polarization-selective_2017}.
In the following article, we show experimentally that this method can improve SHG conversion efficiency by more than an order of magnitude in an $x$-cut resonator.
These results also apply to $z$-cut resonators.


\section{Selective Coupling \label{Theory}}

In order to quantify the selective coupling we introduce the quality factor $Q$.
The $Q$ factors of WGMs are determined by the resonance frequency, $\omega$, in addition to the linewidth, $\Delta\omega$. The linewidth can be attributed to the intrinsic loss rate, $\gamma^{0}$, from absorption in the resonator, and the coupling rate, $\gamma^{c}$, into and out of the resonator.
Each of these partially contributes to the total linewidth, and hence inversely to the $Q$ factor:
\begin{equation}
\frac{1}{Q} = \frac{\Delta\omega}{\omega} = \frac{2(\gamma^{0}+\gamma^{c})}{\omega} = \frac{1}{Q^{0}}+\frac{1}{Q^{c}}.
\end{equation}
Typically, the intrinsic loss rate is fixed (by the bulk dielectric absorption and surface scattering).
The coupling rate, however, is only dependent on the coupling mechanism used; in the case of WGMs, this is evanescent coupling, which has exponential distance dependence \cite{gorodetsky_optical_1999}. 
Therefore, by moving the coupler with respect to the resonator, we can continuously adjust the coupling rate.
This is important, as the maximum intracavity power is reached when $\gamma^{c}=\gamma^{0}$. This condition is known as \emph{critical coupling}.

Critical coupling of the pump, however, does not ensure that the second harmonic is also critically coupled.
The distance, over which the evanescent field decays, is dependent on the mode's wavelength.
In particular, it has been shown \cite{gorodetsky_optical_1999} that the coupling rate to a WGM resonator using prism coupling in air is
\begin{equation}
	\label{eq:coupling_distance}
	\gamma^c \propto \sqrt{\lambda} e^{- 2 \kappa d},\quad\text{where}\quad \kappa\approx\frac{2\pi}{\lambda}\sqrt{n_r^2-1},
\end{equation}
$n_r$ is the effective refractive index for the mode, $d$ is the distance between the prism and resonator, and $\lambda$ is the vacuum wavelength of the mode.
Accordingly, the coupling rates for SHG and pump are---in general---different.
In particular, for high-$Q$ resonances, the intrinsic loss rate is very small, so critical coupling is reached at small coupling rates.
In this case, the exponential term dominates and the pump couples critically at a much farther distance than the signal.

In this next section, we will show---using a simple application of coupled mode theory---that simultaneous critical coupling of both fields is optimal for SHG.
The rate equations for SHG in a cavity can be written as \cite{haus_waves_1984}
\begin{align}
	\frac{dA_p}{dt} &= - \Gamma_p A_p - 2 i K A_p^* A_s + F_p, \\
	\frac{dA_s}{dt} &= - \Gamma_s A_s + i K A_p^2,
\end{align}
where subscript $s$ and $p$ represent signal (second harmonic) and pump modes, $A$ is the slowly varying amplitude of the mode, $\Gamma = \gamma^{c} + \gamma^0 - i \delta \omega$, where $\delta\omega$ is the detuning from the resonance, and $K$ is the second harmonic coupling coefficient. To use coupled mode theory, the amplitudes must be normalized to energy, so naturally we choose $|A|^2=W$, where $W$ is the energy stored in the mode. 
With this normalization, the pump force is $F_p = \sqrt{2 \gamma^c_p} S_p$, where $S_p$ is the pump source amplitude, which satisfies $|S_p|^2=P\nii{incident}$ for the incident pump power. This connects the incoming pump field with the intracavity fields.

In our experiment we assume that the system has reached a steady state, and that we can use the undepleted pump approximation ($2 i K A_p^* A_s \approx 0$), whence we find for the intracavity SHG field (while assuming we are not detuned from the resonances, $\delta\omega_{p,s} = 0$):
\begin{equation}
	A_s = i K \frac{2 \gamma^c_p}{(\gamma^c_p + \gamma^0_p)^2 (\gamma^c_s + \gamma^0_s)} P\nii{incident}.
\end{equation}
The out-coupled second harmonic power is then given by $P\nii{SHG} = |\sqrt{2 \gamma^c_s} A_s|^2$, hence
\begin{equation}
	P\nii{SHG} = {8 K^2} \frac{(\gamma^c_p)^2 \gamma^c_s}{(\gamma^c_p + \gamma^0_p)^4 (\gamma^c_s + \gamma^0_s)^2} P\nii{incident}^2.
\end{equation}
For a given pump power, and fixed intrinsic loss rates, the maximum of this function is reached when both pump and signal are critically coupled ($\gamma^c_{p,s} = \gamma^0_{p,s}$).

\begin{figure}
	\includegraphics[width=3.4in]{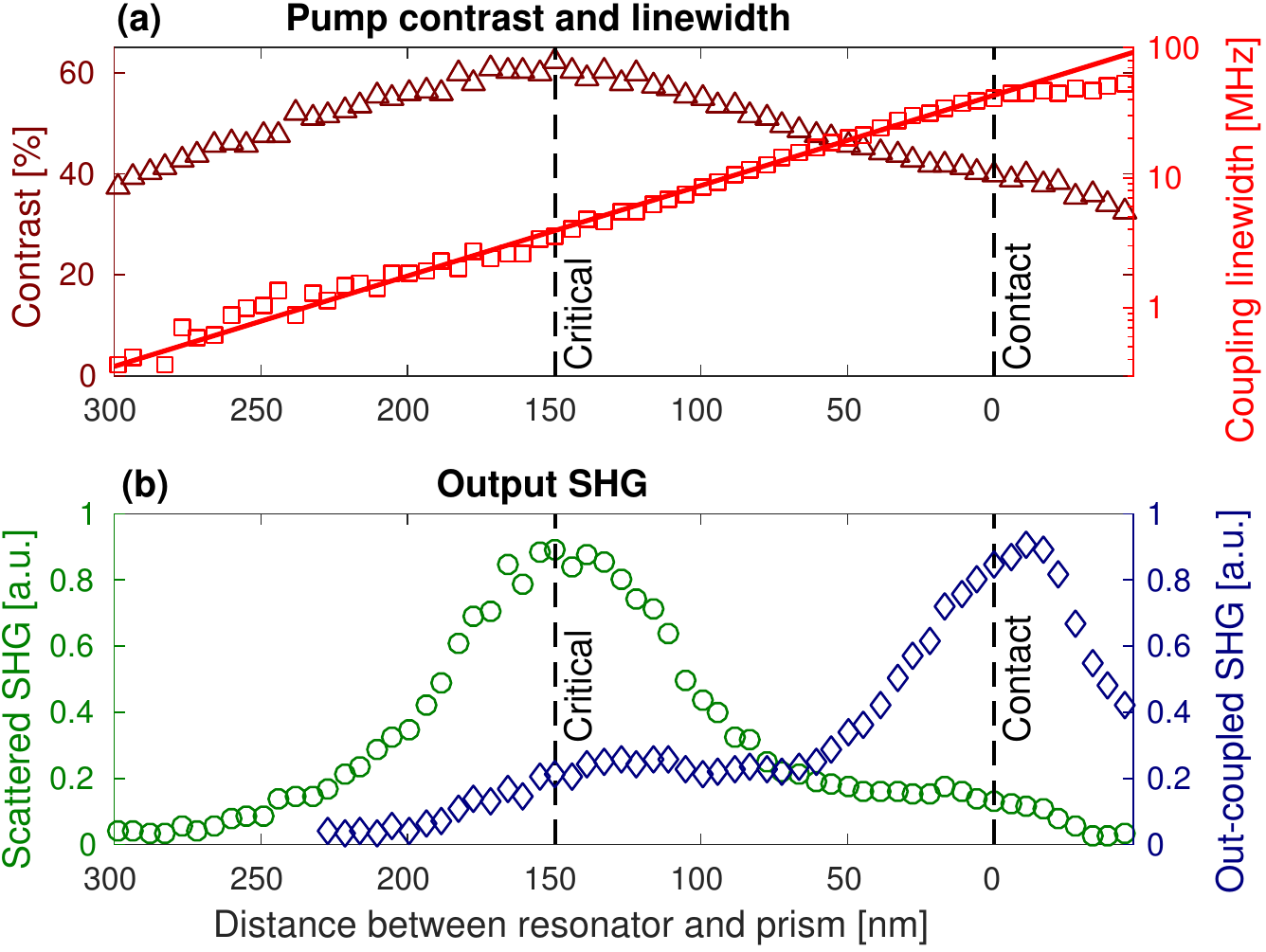}
	\caption{
		\label{fig:scattering}
		SHG as a function of coupler distance.
		In (a), the pump mode's contrast and linewidth are plotted as the prism is moved closer to the resonator till the two are on contact.
		At about \SI{150}{\nano\meter} away, the contrast is maximum and the resonance is critically coupled.
		The behavior of the coupling linewidth changes from exponential at some point.
		We believe this is due to contact between the prism and resonator.
		Meanwhile in (b), we measure the second harmonic both scattered from the resonator, and coupled out via the prism.
		The scattered second harmonic gives an indication of the intracavity power.
		This reaches a maximum when the prism is critically coupled to the pump, but the out-coupled power follows much later when the signal is better coupled at a closer distance.
		This is even when we believe the prism is depressing the resonator.
		The method for this experiment is described in detail in Sec.\,\ref{Scattering}.
	}
\end{figure}

Here there is thus a trade-off, which we illustrate in Fig.\,\ref{fig:scattering}.
To create this figure, a diamond prism is used to couple transverse electric (TE) light of \SI{1548.36}{\nano\meter} into an $x$-cut congruent lithium niobate (LN) WGM resonator.
This prism can be moved by a piezoelectric linear stage closer to or farther from the resonator.
The pump's second harmonic is generated as a TM mode inside the resonator.
To ensure maximum conversion of the pump, it needs to be critically coupled.
However, to obtain the most useful out-coupled second harmonic power, the second harmonic should also be critically coupled.
This occurs when the prism is much closer to the resonator, due both to the increased decay in Eq.\,\eqref{eq:coupling_distance}, and also to the fact that TM modes couple worse in general \cite{foreman_dielectric_2016}.
Could we achieve the best conversion and best signal coupling simultaneously, then we would improve the second-harmonic output.
For a much more detailed description of this experiment, see Sec.\,\ref{Scattering}.

This is where selective coupling comes in.
By adding an additional coupler, which only couples to the second harmonic, we can maintain critical coupling of the pump while \emph{at the same time} critically coupling the second harmonic.

\begin{figure}
	\includegraphics[width=3.4in]{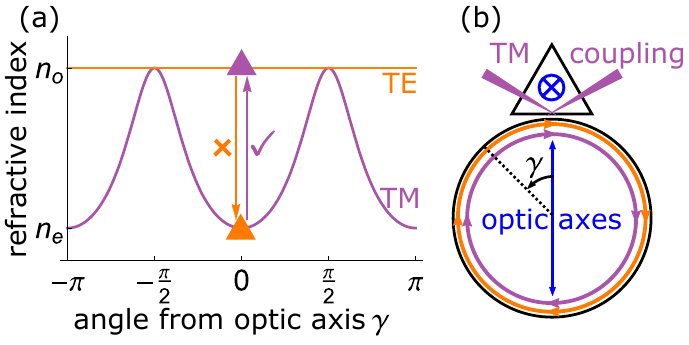}
	\caption{
		\label{fig:selective_in_x}
		Selective coupling in a negative uniaxial, $x$-cut WGM resonator.
		Subfigure (a) shows the refractive indices for TE (orange) and TM (purple) modes at different angles, $\gamma$, from the resonator's optic axis, as well as inside a prism for selective coupling made from the same material.
		This prism is illustrated in (b). It is placed at the resonator's optic axis, and has its optic axis perpendicular to that of the resonator.
		Modes can evanescently couple to a prism if their refractive index inside the prism is greater than or equal to that in the resonator.
		Therefore, only TM modes can couple to this prism.
		In a positive crystal, TE modes would couple at this prism.
	}
\end{figure}

As we have seen in Fig.\,\ref{fig:scattering}, critical coupling of both the signal and pump is normally not possible. 
However, in Fig.\,\ref{fig:selective_in_x} we show how it is possible to selectively couple to TM modes in an $x$-cut scheme for a negative uniaxial crystal.
The TE modes experience the ordinary refractive index, $n_o$, (up to geometric dispersion, which is negligible in our scheme \cite{breunig_whispering_2013}).
Meanwhile, TM modes experience an oscillating refractive index between approximately the ordinary and extraordinary refractive indices, $n_o$ and $n_e$.
At the two points on the resonator, where the the $\mathbf{k}$ vector of the mode is orthogonal to the optic axis, the TM refractive index is purely extraordinary.
Here, the difference in refractive indices for the two modes is at a maximum.
To utilize this difference, we place a $z$-cut prism of the same negative crystal at one of these points.
In this prism, TE and TM modes experience the opposite refractive index to that inside the resonator.
For evanescent coupling to occur, the refractive index in the prism must be greater than that in the resonator.
As the crystal is negative ($n_o>n_e$), TM modes will couple to this prism, whereas TE modes will not.
In positive crystals, TE modes can be selectively coupled in this manner.
In our $x$-cut scheme signal modes are polarized orthogonally to the pump modes.
Normal dispersion dictates that for negative (positive) crystals these signal modes are therefore TM (TE) polarized.
These are precisely the modes we can selectively couple to using the method shown in Fig.\,\ref{fig:selective_in_x}.
We described this scheme in greater detail in Ref.\,\cite{sedlmeir_polarization-selective_2017}.

\section{Experimental Results \label{Experimental}}

In order to experimentally test our hypothesis, we manufactured a \SI[number-unit-product={\text{-}}]{1.2}{\milli\meter} radius resonator from $x$-cut
\singlequote{optical grade} congruent LN from MTI Corporation.
The resonator was initially sanded on a lathe and then polished with diamond slurry in multiple steps from \SI{9}{\micro\meter} down to \SI{0.25}{\micro\meter} grit sizes.
It has an intrinsic $Q$ factor near \SI{1550}{\nano\meter} of $1.7\times 10^8$.

\begin{figure}
	\includegraphics[width=3.4in]{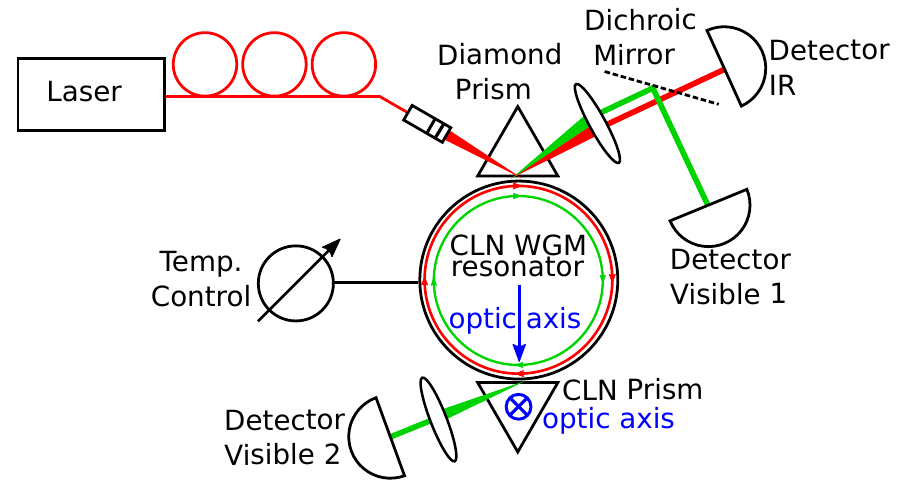}
	\caption{
		\label{fig:setup}
		The experimental setup.
		Red indicates pump ({\raise.20ex\hbox{$\scriptstyle\sim$}}\SI{1550}{\nano\meter}) light.
		Green shows the path of the second harmonic when one is generated.
		Pump light is coupled into an $x$-cut congruent LN WGM resonator using a diamond prism, which can be moved with a piezoelectric linear stage, at the resonator's optic axis.
		Its reflected spectrum is measured with an InGaAs detector.
		When second light harmonic is generated, it can be out-coupled at the diamond prism and measured with a Si detector.
		In addition, there is a $z$-cut LN prism located opposite to the diamond prism controlled by another piezoelectric linear stage.
		Second harmonic (TM-polarized) light can couple out through this prism, but pump (TE-polarized) light cannot.
	}
\end{figure}

The resonator is placed in the experimental setup illustrated in Fig.\,\ref{fig:setup}.
The pump laser is swept in frequency near \SI{1550}{\nano\meter} and coupled to free space using a pigtailed ferrule and graded index (GRIN) lens.
Here it is internally reflected in a diamond prism, with the normal of its coupling surface parallel to the optic axis of the resonator.
The coupling angle is adjusted such that fundamental WGMs are excited efficiently.
This diamond prism sits atop a piezoelectric linear stage (Attocube) for control of the coupling distance on the nanometer scale,
allowing us to adjust the coupling rate.
The reflected light is focused onto an InGaAs detector (Thorlabs), and observed on an oscilloscope.
At the used coupling angle, no TM-polarized pump light can couple into the resonator.
We therefore adjust the polarization using fiber paddles, such that the resonances have the highest contrast.

When a second harmonic is generated inside the resonator, it can be coupled out of the diamond prism, where it is focused, and reflected by a dichroic mirror---separating pump and signal---onto a Si detector (Thorlabs) for observation.
Opposite the diamond prism is a $z$-cut LN prism, which sits atop a second piezoelectric linear stage for independent control of this prism's coupling rate.
As discussed above, only TM-polarized light can couple to this prism.
Light out-coupled there is focused onto a second Si detector (Thorlabs) for observation.

We implement two forms of temperature control. The resonator is surrounded by a resistor-heated temperature-stabilized casing controlled by a PID controller.
This is kept at a stable temperature during the tests, as we would like to keep the resonant frequencies roughly stable.
When the diamond and LN prisms are moved to adjust the coupling rates during the tests, they cause dielectric tuning of the modes \cite{foreman_dielectric_2016}, which changes the phase matching condition and hence inhibits conversion.
To compensate this, we use a \SI{405}{\nano\meter} laser diode as fast temperature control. The diode is fiber-coupled and shone at the resonator from a bare fiber tip.
The up to \SI{5.8}{\milli\watt} of power allows us to tune the pump resonance by about \SI{500}{\mega\hertz} or an equivalent shift to about \SInsp{1.5}{\degreeCelsius} of the resistive heating.
This does not appear to have any negative effect on the resonator, despite congruent lithium niobate's low optical damage threshold \cite{leidinger_impact_2016,savchenkov_photorefractive_2006,savchenkov_photorefractive_2007,savchenkov_enhancement_2006}. 

\subsection{\label{Scattering} Demonstration of different coupling distances}

To show that maximum intracavity SHG---but not maximum out-coupled SHG---is reached when the pump is critically coupled, we use the setup shown in Fig.\,\ref{fig:setup}, but with the modification that the LN prism and its detector are replaced simply by a high-gain detector pointed at the resonator and a focusing lens to measure scattered light.

The laser is swept in frequency over a mode at \SI{1548.36}{\nano\meter}. It is maintained at constant power during this test.
The reflected spectrum obtained during this sweep allows us to fit the contrast and linewidth of the mode.
The pump was initially under-coupled to the point that very little SHG could be measured.
The voltage on the piezoelectric stage was increased in \SI{0.1}{\volt} steps (about \SI{5.5}{\nano\meter}, estimated from the exponential decay of the linewidth \cite{sedlmeir_polarization-selective_2017}), moving the diamond prism closer to the resonator.
At each step, the reflected pump spectrum, as well as out-coupled and scattered second harmonic spectra were recorded.
The pump contrast and linewidth are plotted in Fig.\,\ref{fig:scattering}(a), and the scattered and out-coupled second harmonic powers in Fig.\,\ref{fig:scattering}(b).
As the detector for the scattered second harmonic could only detect a portion of the total scattered light, the emitted powers have been kept in arbitrary units to visualize qualitatively the scaling of powers with coupling.

As can be seen in Fig.\,\ref{fig:scattering}, the pump contrast increases as the prism is moved closer to the resonator until it is about \SI{150}{\nano\meter} away.
Here the pump is critically coupled, and the scattered second harmonic power reaches its maximum.
This shows that the intracavity SHG is at a maximum at that point.
The out-coupled second harmonic, however, is relatively weak until the pump is heavily over-coupled at a much closer distance.
Here, the resonant enhancement of the pump is hindered, causing lower conversion.

\subsection{\label{Selective} Selective coupling}

\begin{figure}
	\includegraphics[width=3.4in]{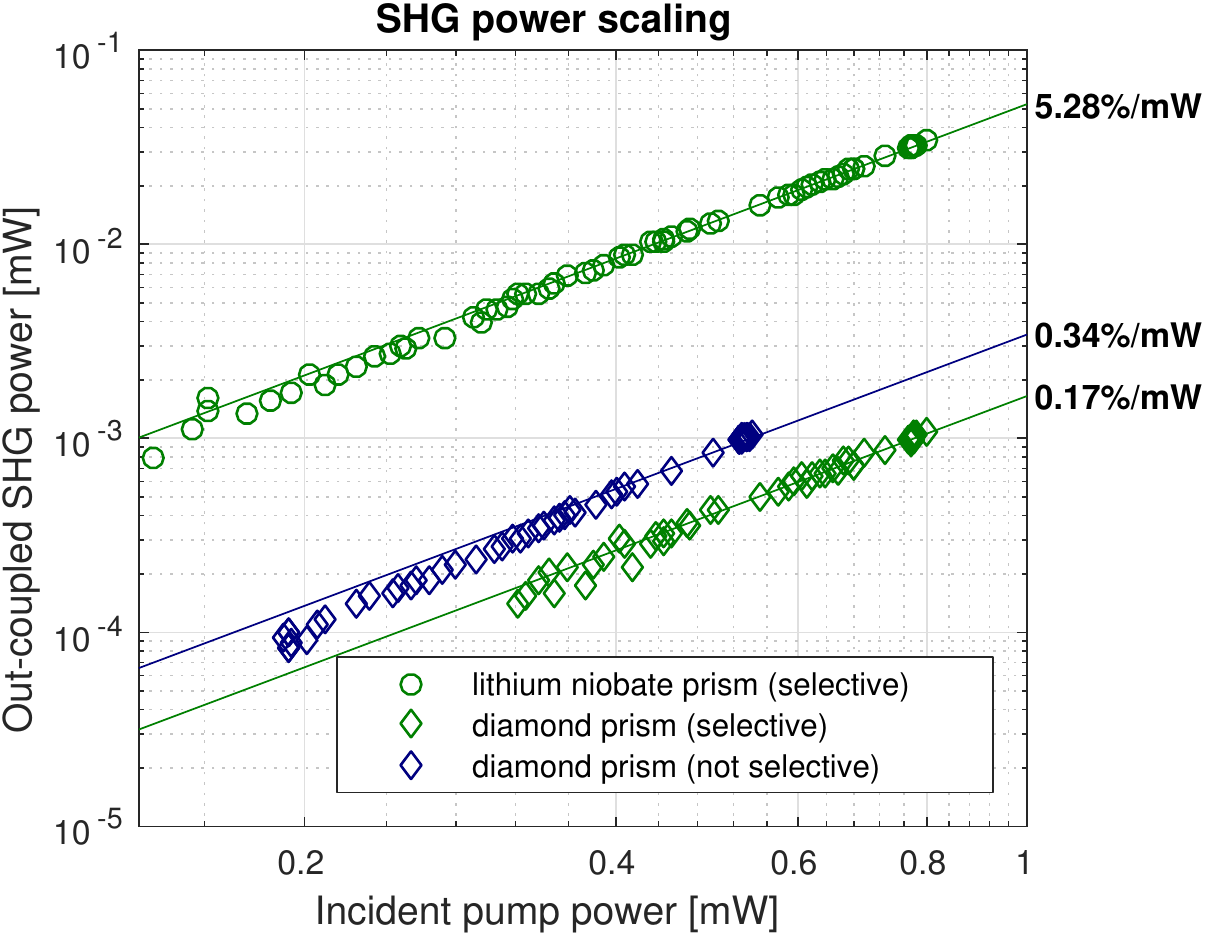}
	\caption{
		\label{fig:selectiveImprove}
		Second harmonic (at \SI{777.7}{\nano\meter}) power as a function of pump power.
		Green markers are measurements where we maximize power selectively out-coupled by the LN prism.
		Blue markers have power out-coupled by the diamond prism maximized (when no LN prism is present).
		The LN prism is represented by circles, and the diamond prism by diamonds.
		The linear data for each curve is fitted to a quadratic, and their intercept at \SI{1}{\milli\watt} is labelled.
		When selective coupling is used, some power couples out of the diamond prism, but this is much less.
		These data were truncated at low SHG power, as their errors approached the noise level.
	}
\end{figure}

To test selective coupling's efficacy, we first selected a conversion channel with high SHG at \SI{1555.4}{\nano\meter}.
During this test, the resistive heating was set to \SInsp{50}{\degreeCelsius}.
The laser was swept in frequency around this mode to ensure the mode volume would not heat up too much due to bulk absorption.
The prisms were moved and the modes tuned by controlling the temperature-tuning laser's power until maximum power was coupled out of the LN prism.
Here the pump power was adjusted to measure how the signal power scaled with pump power.
After these measurements, the LN prism was moved back far enough to ensure no second harmonic would couple out of it.
The diamond prism and tuning laser power were then adjusted to obtain maximum power coupled out of the diamond prism, and again the pump power was adjusted to measure the power scaling.
The results of these measurements are shown in Fig.\,\ref{fig:selectiveImprove}.


Detected signal power clearly increases quadratically with pump power in all measurements.
Power law fits of the data (of the form $P\nii{SHG}\propto P\nii{incident}^\alpha$) give $\alpha$ within 3\% of quadratic for the selectively coupled case.
For the non-selective case, we find a \SInsp{14\pm 9}{\percent} deviation. This could be explained by the modes slightly detuning from each other at lower power.
This quadratic scaling in addition to the wavelength sensitivity range of the Si detectors shows that the signal is the second harmonic.
When compared to the \singlequote{Incident power} shown in Fig.\,\ref{fig:selectiveImprove}, the efficiency of SHG without selective coupling is \SInsp{0.34}{\ppmw}, when the fit is constrained to be quadratic.
When selective coupling is implemented, there is an efficiency of \SInsp{5.28}{\ppmw} at the selective coupling prism, and \SInsp{0.17}{\ppmw} at the diamond prism.
The best possible conversion with the selective coupler is hence fifteen times larger than the best possible conversion with a single diamond prism.

Critical coupling of the pump laser into this mode yielded the maximum contrast of \SInsp{40.7}{\percent}.
It is known that $>\SInsp{99}{\percent}$ prism coupling efficiency is possible by controlling the ratio of the resonator's radii \cite{strekalov_efficient_2009}.
As the resonator rim was sanded, we could not control this ratio, therefore we have scaled the incident powers to be \SInsp{40.7}{\percent} of the laser power.
This scaling is very important for the efficiency measurements for a somewhat subtle reason.
For maximum SHG coupled out of the diamond prism, the pump is over-coupled, meaning that its contrast is lower than \SInsp{40.7}{\percent}.
A typical contrast in this case was a mere \SInsp{9}{\percent}.
In this typical case, with \SI{1}{\milli\watt} laser power, \SI{90}{\micro\watt} would be coupled in, but there would be $\SI{407}{\micro\watt}$ incident pump power, which \emph{could have been} coupled in, had the resonance been critically coupled.
When selective coupling is used, this extra coupling is achieved, therefore the extra power is not discounted.
Failure to account for this would give a conversion efficiency for the non-selective case of \SInsp{6.53}{\ppmw}, because although the second harmonic generated for similar laser powers in that case is fifteen-times less, the contrast is simultaneously more than four-times lower.
This use of in-coupled power when critically coupled is notably different to Refs.\,\cite{lin_wide-range_2013, lin_continuous_2014, furst_second-harmonic_2015}, where unadulterated in-coupled power is used.

\section{Conclusion \label{Conclusion}}

To conclude, we have extended the selective coupling of Ref.\,\cite{sedlmeir_polarization-selective_2017} to uniaxial $x$-cut WGM resonators.
In particular, we experimentally demonstrated this in TM second-harmonic modes in a congruent lithium niobate resonator.
In this scheme, selective coupling is particularly useful for harmonic generation, due to the vastly different coupling distances required for critical coupling.
We showed both theoretically and experimentally that simultaneous critical coupling of signal and pump modes allows for increased second harmonic efficiency; because of selective coupling, we were able to increase this fifteen-fold.
These individually accessible coupling rates will be particularly interesting for applications studying the quantum aspects of nonlinearly generated light in WGMs \cite{fortsch_highly_2015,fortsch_near-infrared_2015,schunk_interfacing_2015,peano_intracavity_2015}.
These results also apply to $z$-cut resonators, when an $x$-cut prism is used.

\begin{acknowledgments}
We acknowledge the extensive support of the Max Planck Institute for the Science of Light, who provided many of the experimental apparatuses used.
\end{acknowledgments}


\begin{thebibliography}{24}%
\makeatletter
\providecommand \@ifxundefined [1]{%
 \@ifx{#1\undefined}
}%
\providecommand \@ifnum [1]{%
 \ifnum #1\expandafter \@firstoftwo
 \else \expandafter \@secondoftwo
 \fi
}%
\providecommand \@ifx [1]{%
 \ifx #1\expandafter \@firstoftwo
 \else \expandafter \@secondoftwo
 \fi
}%
\providecommand \natexlab [1]{#1}%
\providecommand \enquote  [1]{``#1''}%
\providecommand \bibnamefont  [1]{#1}%
\providecommand \bibfnamefont [1]{#1}%
\providecommand \citenamefont [1]{#1}%
\providecommand \href@noop [0]{\@secondoftwo}%
\providecommand \href [0]{\begingroup \@sanitize@url \@href}%
\providecommand \@href[1]{\@@startlink{#1}\@@href}%
\providecommand \@@href[1]{\endgroup#1\@@endlink}%
\providecommand \@sanitize@url [0]{\catcode `\\12\catcode `\$12\catcode
  `\&12\catcode `\#12\catcode `\^12\catcode `\_12\catcode `\%12\relax}%
\providecommand \@@startlink[1]{}%
\providecommand \@@endlink[0]{}%
\providecommand \url  [0]{\begingroup\@sanitize@url \@url }%
\providecommand \@url [1]{\endgroup\@href {#1}{\urlprefix }}%
\providecommand \urlprefix  [0]{URL }%
\providecommand \Eprint [0]{\href }%
\providecommand \doibase [0]{http://dx.doi.org/}%
\providecommand \selectlanguage [0]{\@gobble}%
\providecommand \bibinfo  [0]{\@secondoftwo}%
\providecommand \bibfield  [0]{\@secondoftwo}%
\providecommand \translation [1]{[#1]}%
\providecommand \BibitemOpen [0]{}%
\providecommand \bibitemStop [0]{}%
\providecommand \bibitemNoStop [0]{.\EOS\space}%
\providecommand \EOS [0]{\spacefactor3000\relax}%
\providecommand \BibitemShut  [1]{\csname bibitem#1\endcsname}%
\let\auto@bib@innerbib\@empty
\bibitem [{\citenamefont {Sedlmeir}\ \emph {et~al.}(2017)\citenamefont
  {Sedlmeir}, \citenamefont {Foreman}, \citenamefont {Vogl}, \citenamefont
  {Zeltner}, \citenamefont {Schunk}, \citenamefont {Strekalov}, \citenamefont
  {Marquardt}, \citenamefont {Leuchs},\ and\ \citenamefont
  {Schwefel}}]{sedlmeir_polarization-selective_2017}%
  \BibitemOpen
  \bibfield  {author} {\bibinfo {author} {\bibfnamefont {F.}~\bibnamefont
  {Sedlmeir}}, \bibinfo {author} {\bibfnamefont {M.~R.}\ \bibnamefont
  {Foreman}}, \bibinfo {author} {\bibfnamefont {U.}~\bibnamefont {Vogl}},
  \bibinfo {author} {\bibfnamefont {R.}~\bibnamefont {Zeltner}}, \bibinfo
  {author} {\bibfnamefont {G.}~\bibnamefont {Schunk}}, \bibinfo {author}
  {\bibfnamefont {D.~V.}\ \bibnamefont {Strekalov}}, \bibinfo {author}
  {\bibfnamefont {C.}~\bibnamefont {Marquardt}}, \bibinfo {author}
  {\bibfnamefont {G.}~\bibnamefont {Leuchs}}, \ and\ \bibinfo {author}
  {\bibfnamefont {H.~G.~L.}\ \bibnamefont {Schwefel}},\ }\bibfield  {title}
  {{\selectlanguage {english}\enquote {\bibinfo {title}
  {Polarization-{Selective} {Out}-{Coupling} of {Whispering}-{Gallery}
  {Modes}},}\ }}\href {\doibase 10.1103/PhysRevApplied.7.024029} {\bibfield
  {journal} {\bibinfo  {journal} {Phys. Rev. Applied}\ }\textbf {\bibinfo
  {volume} {7}},\ \bibinfo {pages} {024029} (\bibinfo {year}
  {2017})}\BibitemShut {NoStop}%
\bibitem [{\citenamefont {Franken}\ \emph {et~al.}(1961)\citenamefont
  {Franken}, \citenamefont {Hill}, \citenamefont {Peters},\ and\ \citenamefont
  {Weinreich}}]{franken_generation_1961}%
  \BibitemOpen
  \bibfield  {author} {\bibinfo {author} {\bibfnamefont {P.~A.}\ \bibnamefont
  {Franken}}, \bibinfo {author} {\bibfnamefont {A.~E.}\ \bibnamefont {Hill}},
  \bibinfo {author} {\bibfnamefont {C.~W.}\ \bibnamefont {Peters}}, \ and\
  \bibinfo {author} {\bibfnamefont {G.}~\bibnamefont {Weinreich}},\ }\bibfield
  {title} {{\selectlanguage {english}\enquote {\bibinfo {title} {Generation of
  {Optical} {Harmonics}},}\ }}\href {\doibase 10.1103/PhysRevLett.7.118}
  {\bibfield  {journal} {\bibinfo  {journal} {Phys. Rev. Lett.}\ }\textbf
  {\bibinfo {volume} {7}},\ \bibinfo {pages} {118--119} (\bibinfo {year}
  {1961})}\BibitemShut {NoStop}%
\bibitem [{\citenamefont {Fürst}\ \emph {et~al.}(2010)\citenamefont {Fürst},
  \citenamefont {Strekalov}, \citenamefont {Elser}, \citenamefont {Lassen},
  \citenamefont {Andersen}, \citenamefont {Marquardt},\ and\ \citenamefont
  {Leuchs}}]{furst_naturally_2010}%
  \BibitemOpen
  \bibfield  {author} {\bibinfo {author} {\bibfnamefont {J.~U.}\ \bibnamefont
  {Fürst}}, \bibinfo {author} {\bibfnamefont {D.~V.}\ \bibnamefont
  {Strekalov}}, \bibinfo {author} {\bibfnamefont {D.}~\bibnamefont {Elser}},
  \bibinfo {author} {\bibfnamefont {M.}~\bibnamefont {Lassen}}, \bibinfo
  {author} {\bibfnamefont {U.~L.}\ \bibnamefont {Andersen}}, \bibinfo {author}
  {\bibfnamefont {C.}~\bibnamefont {Marquardt}}, \ and\ \bibinfo {author}
  {\bibfnamefont {G.}~\bibnamefont {Leuchs}},\ }\bibfield  {title}
  {{\selectlanguage {english}\enquote {\bibinfo {title} {Naturally
  {Phase}-{Matched} {Second}-{Harmonic} {Generation} in a
  {Whispering}-{Gallery}-{Mode} {Resonator}},}\ }}\href {\doibase
  10.1103/PhysRevLett.104.153901} {\bibfield  {journal} {\bibinfo  {journal}
  {Phys. Rev. Lett.}\ }\textbf {\bibinfo {volume} {104}},\ \bibinfo {pages}
  {153901} (\bibinfo {year} {2010})}\BibitemShut {NoStop}%
\bibitem [{\citenamefont {Fürst}\ \emph {et~al.}(2015)\citenamefont {Fürst},
  \citenamefont {Buse}, \citenamefont {Breunig}, \citenamefont {Becker},
  \citenamefont {Liebertz},\ and\ \citenamefont
  {Bohatý}}]{furst_second-harmonic_2015}%
  \BibitemOpen
  \bibfield  {author} {\bibinfo {author} {\bibfnamefont {J.~U.}\ \bibnamefont
  {Fürst}}, \bibinfo {author} {\bibfnamefont {K.}~\bibnamefont {Buse}},
  \bibinfo {author} {\bibfnamefont {I.}~\bibnamefont {Breunig}}, \bibinfo
  {author} {\bibfnamefont {P.}~\bibnamefont {Becker}}, \bibinfo {author}
  {\bibfnamefont {J.}~\bibnamefont {Liebertz}}, \ and\ \bibinfo {author}
  {\bibfnamefont {L.}~\bibnamefont {Bohatý}},\ }\bibfield  {title}
  {{\selectlanguage {english}\enquote {\bibinfo {title} {Second-harmonic
  generation of light at 245 nm in a lithium tetraborate whispering gallery
  resonator},}\ }}\href {\doibase 10.1364/OL.40.001932} {\bibfield  {journal}
  {\bibinfo  {journal} {Opt. Lett.}\ }\textbf {\bibinfo {volume} {40}},\
  \bibinfo {pages} {1932--1935} (\bibinfo {year} {2015})}\BibitemShut {NoStop}%
\bibitem [{\citenamefont {Sturman}\ and\ \citenamefont
  {Breunig}(2011)}]{sturman_generic_2011}%
  \BibitemOpen
  \bibfield  {author} {\bibinfo {author} {\bibfnamefont {B.}~\bibnamefont
  {Sturman}}\ and\ \bibinfo {author} {\bibfnamefont {I.}~\bibnamefont
  {Breunig}},\ }\bibfield  {title} {{\selectlanguage {english}\enquote
  {\bibinfo {title} {Generic description of second-order nonlinear phenomena in
  whispering-gallery resonators},}\ }}\href {\doibase 10.1364/JOSAB.28.002465}
  {\bibfield  {journal} {\bibinfo  {journal} {J. Opt. Soc. Am. B}\ }\textbf
  {\bibinfo {volume} {28}},\ \bibinfo {pages} {2465--2471} (\bibinfo {year}
  {2011})}\BibitemShut {NoStop}%
\bibitem [{\citenamefont {Strekalov}\ \emph {et~al.}(2016)\citenamefont
  {Strekalov}, \citenamefont {Marquardt}, \citenamefont {Matsko}, \citenamefont
  {Schwefel},\ and\ \citenamefont {Leuchs}}]{strekalov_nonlinear_2016}%
  \BibitemOpen
  \bibfield  {author} {\bibinfo {author} {\bibfnamefont {D.~V.}\ \bibnamefont
  {Strekalov}}, \bibinfo {author} {\bibfnamefont {C.}~\bibnamefont
  {Marquardt}}, \bibinfo {author} {\bibfnamefont {A.~B.}\ \bibnamefont
  {Matsko}}, \bibinfo {author} {\bibfnamefont {H.~G.~L.}\ \bibnamefont
  {Schwefel}}, \ and\ \bibinfo {author} {\bibfnamefont {G.}~\bibnamefont
  {Leuchs}},\ }\bibfield  {title} {{\selectlanguage {english}\enquote {\bibinfo
  {title} {Nonlinear and quantum optics with whispering gallery resonators},}\
  }}\href {\doibase 10.1088/2040-8978/18/12/123002} {\bibfield  {journal}
  {\bibinfo  {journal} {Journal of Optics}\ }\textbf {\bibinfo {volume} {18}},\
  \bibinfo {pages} {123002} (\bibinfo {year} {2016})}\BibitemShut {NoStop}%
\bibitem [{\citenamefont {Breunig}(2016)}]{breunig_three-wave_2016}%
  \BibitemOpen
  \bibfield  {author} {\bibinfo {author} {\bibfnamefont {I.}~\bibnamefont
  {Breunig}},\ }\bibfield  {title} {{\selectlanguage {english}\enquote
  {\bibinfo {title} {Three-wave mixing in whispering gallery resonators},}\
  }}\href {\doibase 10.1002/lpor.201600038} {\bibfield  {journal} {\bibinfo
  {journal} {Laser \& Photonics Reviews}\ }\textbf {\bibinfo {volume} {10}},\
  \bibinfo {pages} {569--587} (\bibinfo {year} {2016})}\BibitemShut {NoStop}%
\bibitem [{\citenamefont {Beckmann}\ \emph {et~al.}(2011)\citenamefont
  {Beckmann}, \citenamefont {Linnenbank}, \citenamefont {Steigerwald},
  \citenamefont {Sturman}, \citenamefont {Haertle}, \citenamefont {Buse},\ and\
  \citenamefont {Breunig}}]{beckmann_highly_2011}%
  \BibitemOpen
  \bibfield  {author} {\bibinfo {author} {\bibfnamefont {T.}~\bibnamefont
  {Beckmann}}, \bibinfo {author} {\bibfnamefont {H.}~\bibnamefont
  {Linnenbank}}, \bibinfo {author} {\bibfnamefont {H.}~\bibnamefont
  {Steigerwald}}, \bibinfo {author} {\bibfnamefont {B.}~\bibnamefont
  {Sturman}}, \bibinfo {author} {\bibfnamefont {D.}~\bibnamefont {Haertle}},
  \bibinfo {author} {\bibfnamefont {K.}~\bibnamefont {Buse}}, \ and\ \bibinfo
  {author} {\bibfnamefont {I.}~\bibnamefont {Breunig}},\ }\bibfield  {title}
  {{\selectlanguage {english}\enquote {\bibinfo {title} {Highly {Tunable}
  {Low}-{Threshold} {Optical} {Parametric} {Oscillation} in {Radially} {Poled}
  {Whispering} {Gallery} {Resonators}},}\ }}\href {\doibase
  10.1103/PhysRevLett.106.143903} {\bibfield  {journal} {\bibinfo  {journal}
  {Phys. Rev. Lett.}\ }\textbf {\bibinfo {volume} {106}},\ \bibinfo {pages}
  {143903} (\bibinfo {year} {2011})}\BibitemShut {NoStop}%
\bibitem [{\citenamefont {Fürst}\ \emph {et~al.}(2016)\citenamefont {Fürst},
  \citenamefont {Sturman}, \citenamefont {Buse},\ and\ \citenamefont
  {Breunig}}]{furst_whispering_2016}%
  \BibitemOpen
  \bibfield  {author} {\bibinfo {author} {\bibfnamefont {J.~U.}\ \bibnamefont
  {Fürst}}, \bibinfo {author} {\bibfnamefont {B.}~\bibnamefont {Sturman}},
  \bibinfo {author} {\bibfnamefont {K.}~\bibnamefont {Buse}}, \ and\ \bibinfo
  {author} {\bibfnamefont {I.}~\bibnamefont {Breunig}},\ }\bibfield  {title}
  {{\selectlanguage {english}\enquote {\bibinfo {title} {Whispering gallery
  resonators with broken axial symmetry: {Theory} and experiment},}\ }}\href
  {\doibase 10.1364/OE.24.020143} {\bibfield  {journal} {\bibinfo  {journal}
  {Opt. Express}\ }\textbf {\bibinfo {volume} {24}},\ \bibinfo {pages} {20143}
  (\bibinfo {year} {2016})}\BibitemShut {NoStop}%
\bibitem [{\citenamefont {Lin}\ \emph {et~al.}(2013)\citenamefont {Lin},
  \citenamefont {Fürst}, \citenamefont {Strekalov},\ and\ \citenamefont
  {Yu}}]{lin_wide-range_2013}%
  \BibitemOpen
  \bibfield  {author} {\bibinfo {author} {\bibfnamefont {G.}~\bibnamefont
  {Lin}}, \bibinfo {author} {\bibfnamefont {J.~U.}\ \bibnamefont {Fürst}},
  \bibinfo {author} {\bibfnamefont {D.~V.}\ \bibnamefont {Strekalov}}, \ and\
  \bibinfo {author} {\bibfnamefont {N.}~\bibnamefont {Yu}},\ }\bibfield
  {title} {{\selectlanguage {english}\enquote {\bibinfo {title} {Wide-range
  cyclic phase matching and second harmonic generation in whispering gallery
  resonators},}\ }}\href {\doibase 10.1063/1.4827538} {\bibfield  {journal}
  {\bibinfo  {journal} {Appl. Phys. Lett.}\ }\textbf {\bibinfo {volume}
  {103}},\ \bibinfo {pages} {181107} (\bibinfo {year} {2013})}\BibitemShut
  {NoStop}%
\bibitem [{\citenamefont {Lin}\ and\ \citenamefont
  {Yu}(2014)}]{lin_continuous_2014}%
  \BibitemOpen
  \bibfield  {author} {\bibinfo {author} {\bibfnamefont {G.}~\bibnamefont
  {Lin}}\ and\ \bibinfo {author} {\bibfnamefont {N.}~\bibnamefont {Yu}},\
  }\bibfield  {title} {{\selectlanguage {english}\enquote {\bibinfo {title}
  {Continuous tuning of double resonance-enhanced second harmonic generation in
  a dispersive dielectric resonator},}\ }}\href {\doibase 10.1364/OE.22.000557}
  {\bibfield  {journal} {\bibinfo  {journal} {Opt. Express}\ }\textbf {\bibinfo
  {volume} {22}},\ \bibinfo {pages} {557--562} (\bibinfo {year}
  {2014})}\BibitemShut {NoStop}%
\bibitem [{\citenamefont {Gorodetsky}\ and\ \citenamefont
  {Ilchenko}(1999)}]{gorodetsky_optical_1999}%
  \BibitemOpen
  \bibfield  {author} {\bibinfo {author} {\bibfnamefont {M.~L.}\ \bibnamefont
  {Gorodetsky}}\ and\ \bibinfo {author} {\bibfnamefont {V.~S.}\ \bibnamefont
  {Ilchenko}},\ }\bibfield  {title} {{\selectlanguage {english}\enquote
  {\bibinfo {title} {Optical microsphere resonators: optimal coupling to
  high-{Q} whispering-gallery modes},}\ }}\href {\doibase
  10.1364/JOSAB.16.000147} {\bibfield  {journal} {\bibinfo  {journal} {J. Opt.
  Soc. Am. B}\ }\textbf {\bibinfo {volume} {16}},\ \bibinfo {pages} {147}
  (\bibinfo {year} {1999})}\BibitemShut {NoStop}%
\bibitem [{\citenamefont {Haus}(1984)}]{haus_waves_1984}%
  \BibitemOpen
  \bibfield  {author} {\bibinfo {author} {\bibfnamefont {H.~A.}\ \bibnamefont
  {Haus}},\ }\href@noop {} {{\selectlanguage {english}\emph {\bibinfo {title}
  {Waves and {Fields} in {Optoelectronics}}}}}\ (\bibinfo  {publisher}
  {Prentice-Hall},\ \bibinfo {year} {1984})\BibitemShut {NoStop}%
\bibitem [{\citenamefont {Foreman}\ \emph {et~al.}(2016)\citenamefont
  {Foreman}, \citenamefont {Sedlmeir}, \citenamefont {Schwefel},\ and\
  \citenamefont {Leuchs}}]{foreman_dielectric_2016}%
  \BibitemOpen
  \bibfield  {author} {\bibinfo {author} {\bibfnamefont {M.~R.}\ \bibnamefont
  {Foreman}}, \bibinfo {author} {\bibfnamefont {F.}~\bibnamefont {Sedlmeir}},
  \bibinfo {author} {\bibfnamefont {H.~G.~L.}\ \bibnamefont {Schwefel}}, \ and\
  \bibinfo {author} {\bibfnamefont {G.}~\bibnamefont {Leuchs}},\ }\bibfield
  {title} {{\selectlanguage {english}\enquote {\bibinfo {title} {Dielectric
  tuning and coupling of whispering gallery modes using an anisotropic
  prism},}\ }}\href {\doibase 10.1364/JOSAB.33.002177} {\bibfield  {journal}
  {\bibinfo  {journal} {J. Opt. Soc. Am. B}\ }\textbf {\bibinfo {volume}
  {33}},\ \bibinfo {pages} {2177--2195} (\bibinfo {year} {2016})}\BibitemShut
  {NoStop}%
\bibitem [{\citenamefont {Breunig}\ \emph {et~al.}(2013)\citenamefont
  {Breunig}, \citenamefont {Sturman}, \citenamefont {Sedlmeir}, \citenamefont
  {Schwefel},\ and\ \citenamefont {Buse}}]{breunig_whispering_2013}%
  \BibitemOpen
  \bibfield  {author} {\bibinfo {author} {\bibfnamefont {I.}~\bibnamefont
  {Breunig}}, \bibinfo {author} {\bibfnamefont {B.}~\bibnamefont {Sturman}},
  \bibinfo {author} {\bibfnamefont {F.}~\bibnamefont {Sedlmeir}}, \bibinfo
  {author} {\bibfnamefont {H.~G.~L.}\ \bibnamefont {Schwefel}}, \ and\ \bibinfo
  {author} {\bibfnamefont {K.}~\bibnamefont {Buse}},\ }\bibfield  {title}
  {{\selectlanguage {english}\enquote {\bibinfo {title} {Whispering gallery
  modes at the rim of an axisymmetric optical resonator: {Analytical} versus
  numerical description and comparison with experiment},}\ }}\href {\doibase
  10.1364/OE.21.030683} {\bibfield  {journal} {\bibinfo  {journal} {Opt.
  Express}\ }\textbf {\bibinfo {volume} {21}},\ \bibinfo {pages} {30683}
  (\bibinfo {year} {2013})}\BibitemShut {NoStop}%
\bibitem [{\citenamefont {Leidinger}\ \emph {et~al.}(2016)\citenamefont
  {Leidinger}, \citenamefont {Werner}, \citenamefont {Yoshiki}, \citenamefont
  {Buse},\ and\ \citenamefont {Breunig}}]{leidinger_impact_2016}%
  \BibitemOpen
  \bibfield  {author} {\bibinfo {author} {\bibfnamefont {M.}~\bibnamefont
  {Leidinger}}, \bibinfo {author} {\bibfnamefont {C.~S.}\ \bibnamefont
  {Werner}}, \bibinfo {author} {\bibfnamefont {W.}~\bibnamefont {Yoshiki}},
  \bibinfo {author} {\bibfnamefont {K.}~\bibnamefont {Buse}}, \ and\ \bibinfo
  {author} {\bibfnamefont {I.}~\bibnamefont {Breunig}},\ }\bibfield  {title}
  {{\selectlanguage {english}\enquote {\bibinfo {title} {Impact of the
  photorefractive and pyroelectric-electro-optic effect in lithium niobate on
  whispering-gallery modes},}\ }}\href {\doibase 10.1364/OL.41.005474}
  {\bibfield  {journal} {\bibinfo  {journal} {Optics Letters}\ }\textbf
  {\bibinfo {volume} {41}},\ \bibinfo {pages} {5474--5477} (\bibinfo {year}
  {2016})}\BibitemShut {NoStop}%
\bibitem [{\citenamefont {Savchenkov}\ \emph
  {et~al.}(2006{\natexlab{a}})\citenamefont {Savchenkov}, \citenamefont
  {Matsko}, \citenamefont {Strekalov}, \citenamefont {Ilchenko},\ and\
  \citenamefont {Maleki}}]{savchenkov_photorefractive_2006}%
  \BibitemOpen
  \bibfield  {author} {\bibinfo {author} {\bibfnamefont {A.~A.}\ \bibnamefont
  {Savchenkov}}, \bibinfo {author} {\bibfnamefont {A.~B.}\ \bibnamefont
  {Matsko}}, \bibinfo {author} {\bibfnamefont {D.}~\bibnamefont {Strekalov}},
  \bibinfo {author} {\bibfnamefont {V.~S.}\ \bibnamefont {Ilchenko}}, \ and\
  \bibinfo {author} {\bibfnamefont {L.}~\bibnamefont {Maleki}},\ }\bibfield
  {title} {\enquote {\bibinfo {title} {Photorefractive effects in magnesium
  doped lithium niobate whispering gallery mode resonators},}\ }\href {\doibase
  doi:10.1063/1.2212055} {\bibfield  {journal} {\bibinfo  {journal} {Applied
  Physics Letters}\ }\textbf {\bibinfo {volume} {88}},\ \bibinfo {pages}
  {241909--3} (\bibinfo {year} {2006}{\natexlab{a}})}\BibitemShut {NoStop}%
\bibitem [{\citenamefont {Savchenkov}\ \emph {et~al.}(2007)\citenamefont
  {Savchenkov}, \citenamefont {Matsko}, \citenamefont {Strekalov},
  \citenamefont {Ilchenko},\ and\ \citenamefont
  {Maleki}}]{savchenkov_photorefractive_2007}%
  \BibitemOpen
  \bibfield  {author} {\bibinfo {author} {\bibfnamefont {A.~A.}\ \bibnamefont
  {Savchenkov}}, \bibinfo {author} {\bibfnamefont {A.~B.}\ \bibnamefont
  {Matsko}}, \bibinfo {author} {\bibfnamefont {D.}~\bibnamefont {Strekalov}},
  \bibinfo {author} {\bibfnamefont {V.~S.}\ \bibnamefont {Ilchenko}}, \ and\
  \bibinfo {author} {\bibfnamefont {L.}~\bibnamefont {Maleki}},\ }\bibfield
  {title} {\enquote {\bibinfo {title} {Photorefractive damage in whispering
  gallery resonators},}\ }\href {\doibase 10.1016/j.optcom.2006.11.029}
  {\bibfield  {journal} {\bibinfo  {journal} {Optics Communications}\ }\textbf
  {\bibinfo {volume} {272}},\ \bibinfo {pages} {257--262} (\bibinfo {year}
  {2007})}\BibitemShut {NoStop}%
\bibitem [{\citenamefont {Savchenkov}\ \emph
  {et~al.}(2006{\natexlab{b}})\citenamefont {Savchenkov}, \citenamefont
  {Matsko}, \citenamefont {Strekalov}, \citenamefont {Ilchenko},\ and\
  \citenamefont {Maleki}}]{savchenkov_enhancement_2006}%
  \BibitemOpen
  \bibfield  {author} {\bibinfo {author} {\bibfnamefont {A.~A.}\ \bibnamefont
  {Savchenkov}}, \bibinfo {author} {\bibfnamefont {A.~B.}\ \bibnamefont
  {Matsko}}, \bibinfo {author} {\bibfnamefont {D.}~\bibnamefont {Strekalov}},
  \bibinfo {author} {\bibfnamefont {V.~S.}\ \bibnamefont {Ilchenko}}, \ and\
  \bibinfo {author} {\bibfnamefont {L.}~\bibnamefont {Maleki}},\ }\bibfield
  {title} {{\selectlanguage {english}\enquote {\bibinfo {title} {Enhancement of
  photorefraction in whispering gallery mode resonators},}\ }}\href {\doibase
  10.1103/PhysRevB.74.245119} {\bibfield  {journal} {\bibinfo  {journal} {Phys.
  Rev. B}\ }\textbf {\bibinfo {volume} {74}},\ \bibinfo {pages} {245119}
  (\bibinfo {year} {2006}{\natexlab{b}})}\BibitemShut {NoStop}%
\bibitem [{\citenamefont {Strekalov}\ \emph {et~al.}(2009)\citenamefont
  {Strekalov}, \citenamefont {Savchenkov}, \citenamefont {Matsko},\ and\
  \citenamefont {Yu}}]{strekalov_efficient_2009}%
  \BibitemOpen
  \bibfield  {author} {\bibinfo {author} {\bibfnamefont {D.~V.}\ \bibnamefont
  {Strekalov}}, \bibinfo {author} {\bibfnamefont {A.~A.}\ \bibnamefont
  {Savchenkov}}, \bibinfo {author} {\bibfnamefont {A.~B.}\ \bibnamefont
  {Matsko}}, \ and\ \bibinfo {author} {\bibfnamefont {N.}~\bibnamefont {Yu}},\
  }\bibfield  {title} {{\selectlanguage {english}\enquote {\bibinfo {title}
  {Efficient upconversion of subterahertz radiation in a high-{Q} whispering
  gallery resonator},}\ }}\href {\doibase 10.1364/OL.34.000713} {\bibfield
  {journal} {\bibinfo  {journal} {Opt. Lett.}\ }\textbf {\bibinfo {volume}
  {34}},\ \bibinfo {pages} {713--715} (\bibinfo {year} {2009})}\BibitemShut
  {NoStop}%
\bibitem [{\citenamefont {Förtsch}\ \emph
  {et~al.}(2015{\natexlab{a}})\citenamefont {Förtsch}, \citenamefont {Schunk},
  \citenamefont {Fürst}, \citenamefont {Strekalov}, \citenamefont {Gerrits},
  \citenamefont {Stevens}, \citenamefont {Sedlmeir}, \citenamefont {Schwefel},
  \citenamefont {Nam}, \citenamefont {Leuchs},\ and\ \citenamefont
  {Marquardt}}]{fortsch_highly_2015}%
  \BibitemOpen
  \bibfield  {author} {\bibinfo {author} {\bibfnamefont {M.}~\bibnamefont
  {Förtsch}}, \bibinfo {author} {\bibfnamefont {G.}~\bibnamefont {Schunk}},
  \bibinfo {author} {\bibfnamefont {J.~U.}\ \bibnamefont {Fürst}}, \bibinfo
  {author} {\bibfnamefont {D.}~\bibnamefont {Strekalov}}, \bibinfo {author}
  {\bibfnamefont {T.}~\bibnamefont {Gerrits}}, \bibinfo {author} {\bibfnamefont
  {M.~J.}\ \bibnamefont {Stevens}}, \bibinfo {author} {\bibfnamefont
  {F.}~\bibnamefont {Sedlmeir}}, \bibinfo {author} {\bibfnamefont {H.~G.~L.}\
  \bibnamefont {Schwefel}}, \bibinfo {author} {\bibfnamefont {S.~W.}\
  \bibnamefont {Nam}}, \bibinfo {author} {\bibfnamefont {G.}~\bibnamefont
  {Leuchs}}, \ and\ \bibinfo {author} {\bibfnamefont {C.}~\bibnamefont
  {Marquardt}},\ }\bibfield  {title} {\enquote {\bibinfo {title} {Highly
  efficient generation of single-mode photon pairs from a crystalline
  whispering-gallery-mode resonator source},}\ }\href {\doibase
  10.1103/PhysRevA.91.023812} {\bibfield  {journal} {\bibinfo  {journal}
  {Physical Review A}\ }\textbf {\bibinfo {volume} {91}},\ \bibinfo {pages}
  {023812} (\bibinfo {year} {2015}{\natexlab{a}})}\BibitemShut {NoStop}%
\bibitem [{\citenamefont {Förtsch}\ \emph
  {et~al.}(2015{\natexlab{b}})\citenamefont {Förtsch}, \citenamefont
  {Gerrits}, \citenamefont {Stevens}, \citenamefont {Strekalov}, \citenamefont
  {Schunk}, \citenamefont {Fürst}, \citenamefont {Vogl}, \citenamefont
  {Sedlmeir}, \citenamefont {Schwefel}, \citenamefont {Leuchs}, \citenamefont
  {Nam},\ and\ \citenamefont {Marquardt}}]{fortsch_near-infrared_2015}%
  \BibitemOpen
  \bibfield  {author} {\bibinfo {author} {\bibfnamefont {M.}~\bibnamefont
  {Förtsch}}, \bibinfo {author} {\bibfnamefont {T.}~\bibnamefont {Gerrits}},
  \bibinfo {author} {\bibfnamefont {M.~J.}\ \bibnamefont {Stevens}}, \bibinfo
  {author} {\bibfnamefont {D.}~\bibnamefont {Strekalov}}, \bibinfo {author}
  {\bibfnamefont {G.}~\bibnamefont {Schunk}}, \bibinfo {author} {\bibfnamefont
  {J.~U.}\ \bibnamefont {Fürst}}, \bibinfo {author} {\bibfnamefont
  {U.}~\bibnamefont {Vogl}}, \bibinfo {author} {\bibfnamefont {F.}~\bibnamefont
  {Sedlmeir}}, \bibinfo {author} {\bibfnamefont {H.~G.~L.}\ \bibnamefont
  {Schwefel}}, \bibinfo {author} {\bibfnamefont {G.}~\bibnamefont {Leuchs}},
  \bibinfo {author} {\bibfnamefont {S.~W.}\ \bibnamefont {Nam}}, \ and\
  \bibinfo {author} {\bibfnamefont {C.}~\bibnamefont {Marquardt}},\ }\bibfield
  {title} {{\selectlanguage {english}\enquote {\bibinfo {title} {Near-infrared
  single-photon spectroscopy of a whispering gallery mode resonator using
  energy-resolving transition edge sensors},}\ }}\href {\doibase
  10.1088/2040-8978/17/6/065501} {\bibfield  {journal} {\bibinfo  {journal}
  {Journal of Optics}\ }\textbf {\bibinfo {volume} {17}},\ \bibinfo {pages}
  {065501} (\bibinfo {year} {2015}{\natexlab{b}})}\BibitemShut {NoStop}%
\bibitem [{\citenamefont {Schunk}\ \emph {et~al.}(2015)\citenamefont {Schunk},
  \citenamefont {Vogl}, \citenamefont {Strekalov}, \citenamefont {Förtsch},
  \citenamefont {Sedlmeir}, \citenamefont {Schwefel}, \citenamefont {Göbelt},
  \citenamefont {Christiansen}, \citenamefont {Leuchs},\ and\ \citenamefont
  {Marquardt}}]{schunk_interfacing_2015}%
  \BibitemOpen
  \bibfield  {author} {\bibinfo {author} {\bibfnamefont {G.}~\bibnamefont
  {Schunk}}, \bibinfo {author} {\bibfnamefont {U.}~\bibnamefont {Vogl}},
  \bibinfo {author} {\bibfnamefont {D.~V.}\ \bibnamefont {Strekalov}}, \bibinfo
  {author} {\bibfnamefont {M.}~\bibnamefont {Förtsch}}, \bibinfo {author}
  {\bibfnamefont {F.}~\bibnamefont {Sedlmeir}}, \bibinfo {author}
  {\bibfnamefont {H.~G.~L.}\ \bibnamefont {Schwefel}}, \bibinfo {author}
  {\bibfnamefont {M.}~\bibnamefont {Göbelt}}, \bibinfo {author} {\bibfnamefont
  {S.}~\bibnamefont {Christiansen}}, \bibinfo {author} {\bibfnamefont
  {G.}~\bibnamefont {Leuchs}}, \ and\ \bibinfo {author} {\bibfnamefont
  {C.}~\bibnamefont {Marquardt}},\ }\bibfield  {title} {{\selectlanguage
  {english}\enquote {\bibinfo {title} {Interfacing transitions of different
  alkali atoms and telecom bands using one narrowband photon pair source},}\
  }}\href {\doibase 10.1364/OPTICA.2.000773} {\bibfield  {journal} {\bibinfo
  {journal} {Optica}\ }\textbf {\bibinfo {volume} {2}},\ \bibinfo {pages}
  {773--778} (\bibinfo {year} {2015})}\BibitemShut {NoStop}%
\bibitem [{\citenamefont {Peano}\ \emph {et~al.}(2015)\citenamefont {Peano},
  \citenamefont {Schwefel}, \citenamefont {Marquardt},\ and\ \citenamefont
  {Marquardt}}]{peano_intracavity_2015}%
  \BibitemOpen
  \bibfield  {author} {\bibinfo {author} {\bibfnamefont {V.}~\bibnamefont
  {Peano}}, \bibinfo {author} {\bibfnamefont {H.~G.~L.}\ \bibnamefont
  {Schwefel}}, \bibinfo {author} {\bibfnamefont {C.}~\bibnamefont {Marquardt}},
  \ and\ \bibinfo {author} {\bibfnamefont {F.}~\bibnamefont {Marquardt}},\
  }\bibfield  {title} {\enquote {\bibinfo {title} {Intracavity {Squeezing}
  {Can} {Enhance} {Quantum}-{Limited} {Optomechanical} {Position} {Detection}
  through {Deamplification}},}\ }\href {\doibase
  10.1103/PhysRevLett.115.243603} {\bibfield  {journal} {\bibinfo  {journal}
  {Physical Review Letters}\ }\textbf {\bibinfo {volume} {115}},\ \bibinfo
  {pages} {243603} (\bibinfo {year} {2015})}\BibitemShut {NoStop}%
\end{thebibliography}
%

\end{document}